\documentstyle[proceedings,numreferences]{crckapb}
\begin{opening}
\title{Cosmochemistry in the early Universe}
\author{Denis Puy}
\institute{Institute of Theoretical Physics, University of Zurich\\
Winterthurerstrasse, 190 - 8057 Zurich (Switzerland)
\\
and
\\
Paul Scherrer Institute, Laboratory for Astrophysics\\
5232 Villigen (Switzerland)}
\end{opening}

\begin{document}

\begin{abstract}
\footnote{Contribution to the 6$^{th}$ Trieste Conference Proceedings,  
{\it First steps in the origin of life in the Universe}, Trieste-Italy, 
18-22 Sptember 2000}
At early times the Universe was filled up with an extremely dense and hot gas. 
Due to the expansion it cooled below the binding energies of atoms which led 
to the formation of the first nuclei. In the physical environment of the 
post-recombination period of hydrogen, molecules such as $H_2$, $HD$ and $LiH$ 
can be formed. The proto-object formation, resulting from the growth of linear 
density fluctuations in the early Universe, can have an 
important impact on the chemical state of the Universe. Hence it can be 
enriched with metals, and thus lead to the formation of the first pre-biotic 
molecules.
\\
In this contribution, I will present some scheme for the formation of 
primordial molecules and discuss the consequence of the formation of 
first stars on the existence of possible primordial pre-biotic.
\end{abstract}

\section{Introduction}
The conventional model of nucleosynthesis is the process by which chemical 
elements and their isotopes are formed, the heavy elements (carbon, oxygen, 
nitrogen and heavier ones) are thought to be the result of thermonuclear 
burning in stars, 
and especially the relatively rare stars which become supernovae 
\cite{signore99}, \cite{signore00}.
\\
Standard big bang nucleosynthesis generated just a few elements: 
hydrogen, deuterium, helium and lithium, traces of 
beryllium and boron. The recombination of this primeval plasma led 
progressively to form the neutral species at $\sim$ 10$^9$ yr 
after the big bang. Little is known about the history of the Universe 
corresponding to 10$^6$-10$^9$ years (or to redshifts of $z\sim 1000-5$) after 
the big bang, and only recently astrophysicists begun to investigate 
seriously the post-recombination era and particularly the problem of the 
formation of the first molecules \cite{lepp84}, 
\cite{puy93}, \cite{galli98}. 
\\
It is usually supposed that the first objects (first stars or galaxies) 
could be formed from a gas cloud by the gravitational attraction. The clouds 
contract adiabatically and some kind of {\it cryogen} is necessary to form 
bound systems. Since the primeval gas does not contain heavy elements like 
carbon, primordial molecules such as $H_2$, $HD$ and $LiH$ are good 
candidates to lead to an efficient 
cooling \cite{puy96}, \cite{puy97}. In this context primordial 
molecules can play an important role in the pregalactic gas 
particularly on the thermochemical stability \cite{puy98a}, \cite{puy98b}.
\\
In this contribution, I will present in Sect. 2 the formation of 
primordial molecules. Then in Sect. 3, I will describe how these molecules 
can influence on the 
formation of the first objects. In this context, I will discuss the important 
influence of these proto-objects on the second generation of molecular 
formation, which will be important for the formation 
of biomolecules. I will describe, in Sect. 4, some possible outlooks of this 
study.

\section{Formation of primordial molecules}
At the end of the recombination period of hydrogen, the ions became 
progressively neutralized in the order of their ionization potentials. 
With the creation of neutral species, molecular ions can initiate the 
molecular formation. The chemistry of the early Universe is 
the chemistry of the elements hydrogen and its isotope deuterium, helium, 
lithium and its isotopic forms. The ongoing physical reactions are immense 
after the 
recombination of hydrogen, the main processes are collisional (ionization, 
radiative recombination, attachement...) and radiative due to the presence of 
the cosmic microwave background radiation (photoionization, 
photodissociation, photodetachement) \cite{puy93}, \cite{galli98}.
\\
Two main routes lead to $H_2$:
\begin{equation}
H^- + H \longrightarrow H_2 + e^- \ \ {\rm and} \ \ 
H_2^+ + H \longrightarrow H_2 + H^+ ,
\end{equation}
when $HD$ is formed through $H_2$ \cite{palla95} and $LiH$ from $LiH^+$ 
\cite{stancil96}
\begin{equation}
H_2 + D^+ \longrightarrow HD + H^+ \ , \ \
LiH^+ + H \longrightarrow LiH + H^+ .
\end{equation}
\noindent
The reaction rates depend on the temperature. Thus the temperature and density 
evolution equations must be solved simultaneously, which needs in turn 
the simultaneous determination of molecular 
cooling and heating rates \cite{puy93}, \cite{galli98}. From the initial 
abundances given by the standard 
big bang nucleosynthesis \cite{sarkar96}, the integration 
of the chemical network lead to the following abundances of the main 
molecules at $z\sim 5$:

\begin{table*}[h!tbp]
\renewcommand{\arraystretch}{1.0}
\centering
\begin{tabular}{||c|c|c|c||} \hline 
$e^-/H$ & $H_2/H$ & $HD/H$ & $LiH/H$\\ \hline
$\sim 3 \times 10^{-4}$ & $\sim 10^{-6}$ & $1.2 \times 10^{-9}$ & 
$\sim 7 \times 10^{-20}$ \\ \hline
\end{tabular}
    \label{tab:molec}
\end{table*}

\section{First stars and cosmochemistry}
The first objects formed after the molecular formation had no metals heavier 
than lithium. 
The role of molecular cooling in the fragmentation of primordial gas 
clouds is important \cite{silk77}. The excitation of the 
rotational levels of $H_2$ and $HD$ induce a 
cooling which can trigger thermal instability in the collapsing cloud 
\cite{puy96}, \cite{puy98b}. Thus, 
primordial molecules play a crucial role on the evolution of the first 
objects, through the process of fragmentation. The first objects can be, in 
some cases, massive stars \cite{tegmark97}.
\\
Thus, as soon as stellar processes occur, proto-objects can lead through SN 
explosions to the formation of other molecules species, such as $CO$, $CI$ or 
$HCN$. They in turn are important sources of 
contamination of the medium, and thus can offer different ways of early 
pre-biotic molecular formation. Chakrabarti showed that 
a significant amount of adenine, a DNA base, may be produced during 
molecular cloud collapse \cite{chakra00}, through the chain reaction ($HCN$ 
addition):
\begin{equation}
HCN + HCN \longrightarrow ...+ HCN ... \longrightarrow
 H_5C_5N_5 \ ({\rm adenine}).
\end{equation}
\noindent
Actually, the direct detection of the prebiotic 
molecules is not easy. For example, the 
tentative of detection of glycine were negative or below the confusion 
limit \cite{combes96}.  

\section{Outlooks}
We have seen that through the formation and the evolution 
of the first objects, it is useful to consider DNA bases that 
have been produced at an early stage of the Universe. 
A large lepton asymmetry in the standard big bang nucleosynthesis 
could lead to the formation of a non-negligible number of 
heavier elements \cite{dolgov00}, 
and thus a strong incidence on the cosmochemistry. For example, 
the existence of primordial 
carbon or nitrogen can produce  primordial bio-molecules or 
pre-biotic molecules such as amino acids, sugars, adenine just after the cosmological 
recombination period.  
\\
Thus, the formation of DNA bases could happen in the early history of 
the Universe. Barrow pointed out the link 
between the biochemical timescales for the evolution of life and the 
astrophysical timescales that determine the time required to create an 
environment supported by a stable hydrogen burning star \cite{barrow98}. 
In the context of pre-biotic molecular 
formation in the early Universe, it is not necessary to establish this natural 
link between the stars and the prebiotic chemistry. Pre-biotic molecules 
could have contamined the first objects and planets from the beginning. 
Nevertheless these molecules could have been destroyed during the 
formation of planetary disks. Comets carrying away these molecules are thus 
interesting 
sites of protection and potential sources of contamination for planetary 
systems.
\\
Thus, life bases could be appeared earlier than what we think. In this context 
the emergence of extraterrestrial intelligence should be not exceedingly rare 
(see \cite{carter83}, \cite{barrow86}, \cite{livio99})
 
\section*{Acknowledgments}

I would like to thank Prof. Chela-Flores for the organization of this 6$^{th}$ 
Trieste Conference, Lukas Grenacher and Flavia Puy-Svara for useful 
discussions and comments. 
Part of this work was supported by the {\it Tomalla Foundation} and by the 
Swiss National Science Foundation.

\end{document}